\documentclass[aps,twocolumn,groupaddress,showpacs,color]{revtex4-1}
\usepackage{graphicx}
\usepackage{amssymb}
\usepackage{color}
\usepackage{amsmath}
\usepackage{textcomp}
\usepackage{amsfonts}
\usepackage{hyperref}

\begin{document}

\title{Synchronization in dynamical networks of locally coupled self-propelled oscillators}

\author{Demian Levis} 
\affiliation{Departament de F\'isica de la Mat\`eria Condensada, Universitat de Barcelona, Mart\'i i Franqu\`es, E08028 Barcelona, Spain}
\author{Ignacio Pagonabarraga} 
\affiliation{Departament de F\'isica de la Mat\`eria Condensada, Universitat de Barcelona, Mart\'i i Franqu\`es, E08028 Barcelona, Spain}
\author{Albert D\'iaz-Guilera} 
\affiliation{Departament de F\'isica de la Mat\`eria Condensada, Universitat de Barcelona, Mart\'i i Franqu\`es, E08028 Barcelona, Spain}
 

\begin{abstract}
Systems of mobile physical entities exchanging information with their neighborhood can be found in many different situations. The understanding of their emergent cooperative behaviour has become  an important issue across disciplines, requiring a general conceptual framework in order to harvest the potential of these systems.  
We study the synchronization of  coupled oscillators in time-evolving networks defined by the positions of self-propelled agents interacting in real space. 
In order to understand the impact of mobility in the synchronization process on general grounds, we introduce a simple model of self-propelled hard disks performing  persistent random walks in 2$d$ space and carrying an internal Kuramoto phase oscillator. 
For non-interacting particles, self-propulsion accelerates synchronization. The competition between agent mobility and excluded volume interactions gives rise to a richer scenario, leading to an optimal self-propulsion speed.
We identify two extreme dynamic regimes where  synchronization  can be understood from theoretical considerations. 
A systematic analysis of our model quantifies the departure from the latter ideal situations and characterizes the different  mechanisms leading the evolution of the system.  
We show that the  synchronization of locally coupled mobile oscillators generically proceeds through  coarsening verifying dynamic scaling and sharing strong similarities with the phase ordering dynamics of the 2$d$ XY model following a quench.
Our results shed light into the generic mechanisms leading the synchronization of mobile agents, providing a efficient way to understand more complex or specific situations involving time-dependent networks where synchronization, mobility and excluded volume are at play.

\end{abstract}

 

\maketitle

\setlength{\textfloatsep}{10pt} 
\setlength{\intextsep}{10pt}

 
\section{Introduction}\label{sec:intro}

Synchronization processes by which a large population of units spontaneously organize into a cooperative state play an important role in very diverse contexts, from physics to biology going through such disparate fields as ecology, sociology or neurosciences, among others~\cite{Pikovsky2003,StrogatzBook}. 
Fireflies flashing at unison or peacemaker cells firing at a given rate are examples of synchronized states where order in time emerges without any centralized control. 
A major breakthrough in the description of such collective phenomena was the Kuramoto model of phase coupled oscillators, which, since its introduction in the mid 1970s,  has become a paradigmatic model in the study of synchronization~\cite{Kuramoto2012, Acebron2005}.
In its original version, each oscillator was considered to be equally coupled to all the others. However, very few systems justify such simplification and several extensions of the Kuramoto model to adapt it to more realistic situations have been introduced since then. 
The synchronization of locally coupled phase oscillators arranged in lattices or more complex network topologies has been widely studied over the past decades, taking advantage of the recent developments on the theory of complex networks~\cite{Arenas2008}.     
To date, most studies of synchronization involve static networks with a fixed topology, despite that many interesting synchronization phenomena involves mobile agents. In this work we turn our attention into the study of time evolving networks, where the interplay between the motion of the agents and the dynamics of their phases plays a key role in the synchronization process.

Large groups of living organisms can cooperate to form complex dynamical patterns across a broad range of length scales, from the macroscale - like a flock of birds protecting the group from a predator~\cite{SumpterBook} - down to the microscale - like a suspension of cells synchronizing their genetic clocks to perform some biological function~\cite{Stricker2008,Danino2010,Prindle2012,Prindle2014,Uriu2014Rev}.
The question of whether it is possible to recreate (and control) emergent group behaviour in artificial systems made of autonomous robots has been recently raised, attracting the attention of a growing number of multidisciplinary researchers interested in self-organisation and swarm behaviour~\cite{Rubenstein2014, Mijalkov2015}. 
At the microscale, interesting examples of synchronization in dynamical networks can be found in the realm of microbiology. 
Recent advances in synthetic biology allow to design  genetic oscillators and monitor their cycles~\cite{Novak2008Rev}. For instance, in the experiments done by Hasty and collaborators, a genetic oscillator that drives the expression of a fluorescent protein was inserted into \emph{E. Coli} bacteria, making them flash at a regular rate~\cite{Stricker2008}. Once coupled to the quorum sensing machinery,  a large population of bacteria can synchronize, flashing all together at the same rhythm~\cite{Danino2010,Prindle2014}.
Almost all organisms use internal clocks to synchronize their behaviour with their environment and perform  physiological tasks in response. It has been shown that during embryonic development, the coordinated expression of genetic oscillators, the so-called segmentation clock, plays a key role in somitogenesis~\cite{Jiang}. This non-exhaustive list of examples shows that the ability to manipulate and control the synchronization of systems made of mobile entities could be exploited as design principles to build biological sensors, bio-inspired materials or improve mobile communication systems. 



Over the last years, it has become usual in the physics community to consider as active matter any system made of objects which are able to self-propel,  locally converting energy from their environment into motion~\cite{Vicsek2012,MarchettiReview}. Animal groups, bacterial suspensions or collections of robots fall into this novel class of so-called 'active soft materials'.  
The key feature of these systems is that the local driving needed to sustain self-propulsion breaks detailed balance, which automatically drives the system far from thermodynamic equilibrium. As a consequence, the competition between interparticle interactions and activity gives rise to a plethora of genuine non-equilibrium phenomena, like the formation of large clusters of particles in the absence of any attraction between them~\cite{BialkeReview}. 
In order to study the impact of mobility in the synchronization of locally coupled phase oscillators, we develop a new model in which agents perform a persistent, or 'active', random walk in two dimensions.
Here we wish to adopt a minimalist approach, neglecting all the bio-chemical and physical details except the ones that we think are truly essential: self-propulsion, steric repulsion and local phase coupling. 
Each agent is modeled as a physical hard particle carrying an internal phase oscillator coupled with its surrounding within a given range. 
We use the model introduced in~\cite{Levis2014} to describe the dynamics of the particles, while the dynamics of the oscillators are based on the Kuramoto model in the time-evolving network defined by their positions. 

Several different models have been recently introduced to study the synchronization of mobile agents. The first model to consider such problem was proposed in~\cite{Bollt2004}. In this work,  the agents positions (nodes) are described by an ergodic, yet deterministic, map and their internal state is described by a chaotic oscillator which is coupled (linked) to those agents that are close enough. In~\cite{Fujiwara2011}, one of the authors has considered point-like Brownian agents moving in two dimensions at low densities and aligning their phases within a given interaction range. 
Motivated by the synchronization of segmentation clocks during embryonic development, Uriu and coworkers introduced a lattice model in two-dimensions where cells exchange their location at a given rate~\cite{Uriu2010, Uriu2012}. The role played by self-propulsion and velocity alignment in the synchronization of cell tissues was also investigated recently~\cite{Uriu2014}. In this latter work, the authors introduce a model of  coupled phase oscillators attached to repulsive spherical particles at very high density conditions close to  jamming. 

Our model is a generalization of these latter stochastic models~\cite{Fujiwara2011,Uriu2010, Uriu2012,Uriu2014}. In the limit of Brownian point-like agents, our model reduces to the one studied in~\cite{Fujiwara2011}, while by increasing the density of hard disk agents we approach the closed packed limit considered in~\cite{Uriu2010, Uriu2012,Uriu2014}. In this work we will establish a more general framework that allows us to explore synchronization in mobile systems at intermediate densities, i.e. in between the two limiting cases considered in the literature until now.     
In the active matter community, it is now quite well known that, below the high density limit, the competition between interactions and activity gives rise to strong spatio-temporal heterogeneities~\cite{Fily2014,Levis2014,BialkeReview}. Our approach allows us to explore the impact of these dynamical structures, generic in active systems, on the synchronization of phase oscillators.    

The article is organized as follow. In section~\ref{sec:model} we present our model of self-propelled coupled oscillators. The details of the numerical simulations we performed to explore the model are given in section~\ref{sec:numerics}. In section~\ref{sec:regimes} we describe  two limit dynamical regimes that we will use as a reference. The deviations from these idealized cases will be discussed in sections~\ref{sec:points} and \ref{sec:disks} where our main results are presented. The dynamics of self-propelled oscillators carried by non-interacting point-like agents is analyzed in section~\ref{sec:points}. In section~\ref{sec:disks} we turn into our general model and explore the impact of mobility and steric effects in the synchronization of locally coupled oscillators carried by self-propelled hard disks. We finally conclude our work in section~\ref{sec:conclusion}.   


\section{Self-propelled  oscillator model}\label{sec:model}

\subsection{Self-propelled particles}\label{sec:SPP}
We consider a population of $N$ moving individuals in two dimensions. In our model, the individuals (or agents) are self-propelled hard disks of diameter $\sigma$ moving off-lattice in a $L\times L$ surface with periodic boundary conditions. 
Each particle $i$, located at $\boldsymbol{r}_i(t)=(x_i(t),y_i(t))$ at time $t$, moves accordingly to the kinetic Monte Carlo model described in detail in~\cite{Levis2014}. 
It is an extension of the Monte Carlo dynamics used to simulate hard disks in which we simply  introduce correlations between successive displacements.
The evolution of the $i$-th particle position is given by 
\begin{equation}\label{eq:Motion01}
 \boldsymbol{r}_i(t+\Delta t)=\boldsymbol{r}_i(t)+\boldsymbol{\delta}_i(t)P_{\text{acc}}\Delta t
\end{equation}
where $P_{\text{acc}}$ is the acceptance probability of the update, which encodes the interaction between particles. Here we use the Metropolis scheme: for a hardcore repulsion,    $P_{\text{acc}} = 1$ if the move does not generate any overlap with a neighbouring disk and  $P_{\text{acc}} = 0$  otherwise. 
Self-propulsion is introduced through the statistical properties of the displacement field $\boldsymbol{\delta}_i$ which, at a given time step $t$,  is constructed as
\begin{equation}\label{eq:Motion02}
 \boldsymbol{\delta}_i(t)=\boldsymbol{\delta}_i(t-\Delta t)+v_1 \boldsymbol{\eta}_i(t)\,,\ \boldsymbol{\delta}_i(t=0)=v_0 \boldsymbol{\eta}_i(0)\,,
\end{equation}
where $\boldsymbol{\eta}_i$ is a random vector with components independently
drawn at each step from a uniform distribution in the interval $[-1,1]$.
In other words, the displacement of a particle at some time $t$ is given by the displacement at the previous time step plus a uniform random shift of typical amplitude $\approx v_1$. 
For $v_1\ll v_0$ the random shift is negligeable and particles move ballistically with velocity $v_0$. 
When the contribution from the random shift becomes larger than the one from the previous step, meaning $v_1\gg v_0$, the displacements are not time-correlated anymore and we recover the Monte Carlo dynamics of equilibrium Brownian disks. In between, an isolated particle  describes an overdamped persistent random walk with \emph{persistence time}  $\tau=(v_0/v_1)^2\Delta t$ (see ~\cite{Levis2014} for further details). 

This model introduces two control parameters: the \emph{packing fraction} $\phi=\pi\sigma^2\rho/4$ (where $\rho=N/V$ is the number density) and the \emph{persistence time} $\tau$, which quantifies the amount of activity in the system and therefore the departure from equilibrium. 
The competition between steric interactions and self-propulsion in this model leads to a rich phase behaviour in the $(\phi-\tau)$ plane, with a fluid, cluster and gel-like phase~\cite{Levis2014}. Despite its simplicity, this model of active disks has been successful in describing the experimental equations of state of suspensions of self-propelled colloids~\cite{Ginot2015}. 

In the absence of excluded volume interactions, $P_{\text{acc}}=1$, the only control parameter in the model is $\tau$. In this limit, that we will study in section~\ref{sec:points}, particles move ballistically at times $t \ll \tau$ and diffusively at longer times $t\gg \tau$. The persistence time sets the crossover between these two regimes.  
In the ballistic regime, the mean square displacement
\begin{equation}
\Delta r^{2}(t)=N^{-1}\sum_i\langle (\boldsymbol{r}_i(0)-\boldsymbol{r}_i(t))^2 \rangle \,,
\end{equation} 
behaves as $\Delta r^{2}(t)=(v_0\,t)^2$, while in the diffusive one $\Delta r^{2}(t)=4Dt$ with $D\propto v_0^2\tau$ (see~\cite{Levis2014} for more details).

\subsection{Phase oscillators}\label{sec:Kuramoto}

Each self-propelled particle possesses an internal phase oscillator denoted $\theta_i$. Its dynamics is described by the Kuramoto model on the dynamical network  defined by the particles' positions (nodes), connected to each other within a prescribed finite range~\cite{Arenas2008}. We consider the situation in which all the oscillators have the same intrinsic frequency which can be taken as zero without loss of generality. This can be formally written as
\begin{equation}\label{eq:Kuramoto} 
\dot{{\theta}}_i(t)={\frac{K}{R_{\theta}^2}}\sum_{j=1}^N  A_{ij} \sin(\theta_j(t)-\theta_i(t)) \, . 
\end{equation}
The \emph{connectivity matrix} $A_{ij}$ determines whether two oscillators $i$ and $j$ interact. Here we use a local connectivity scheme based on the two-dimensional Euclidean distance between oscillators:
 $A_{ij}=1$ if $|\boldsymbol{r}_i-\boldsymbol{r}_j|\leq R_{\theta}$ and $A_{ij}=0$ otherwise. This means that two oscillators  interact if their centers are separated by a distance shorter than the \emph{interaction radius} $R_{\theta}$. Thus, the adjacency matrix defining the network structure where the Kuramoto model is defined is time dependent. 
 The \emph{coupling strength} between any pair of oscillators in the interaction range is controlled by $K$.
For $K>0$, oscillators at a distance smaller than $R_{\theta}$ approach their phases. 
Note that excluded volume interactions impose that $R_{\theta}>\sigma$ in order to have any phase coupling at all. This model of coupled phase oscillators introduces two control parameters into the problem; $R_{\theta}$,  which controls the static local topology of the interaction network, and $K$, which quantifies the tendency of the oscillators to synchronize their phases.

In the absence of motion, our model is equivalent to the Kuramoto model in a (static) random geometric network~\cite{Arenas2008}.  
For $1/\sqrt{\rho}\gtrsim R_{\theta}$, the interaction network is not simply connected (a geometrical percolation transition occurs at $R^p_{\theta}(\rho)\approx\sqrt{4.51/\rho\pi}$~\cite{Dall2002}). Disconnected components can locally synchronize but global synchronization across the whole system cannot be reached. 
However, in our model,  the presence of motion makes the topology of the network dynamic:  the connectivity matrix $A_{ij}(t)$ becomes time-dependent, allowing originally disconnected oscillators at $t=0$ to interact for some period of time. Then, thanks to mobility, the system can globally synchronize even at low densities~\cite{Bollt2004,Bollt2006,Fujiwara2011}.

In the limit $R_{\theta}\approx L$ the interactions are all-to-all - $A_{ij}=1$, $\forall$ $i,j$ - and, in the absence of motion, our system reduces to the mean-field model introduced by Kuramoto with a delta-distribution of intrinsic frequencies~\cite{Kuramoto2012}. In the original Kuramoto model, each oscillator evolves with a natural frequency $\omega_i$ drawn from a unimodal distribution $g(\omega)$. The notion of distance is lost in this mean-field model and the network becomes fully connected, allowing an analytic treatment~\cite{Acebron2005}.  A synchronized state, characterised by a finite fraction of oscillators sharing the same phase, emerges for $K>K_c$. The critical coupling $K_c$ is proportional to the variance of the distribution $g(\omega)$, which on our case is strickly zero. 

To summarize the main features of our model, the motion of the units defined by the set of eqs.~(\ref{eq:Motion01}) and (\ref{eq:Motion02}) takes into account the competition between self-propulsion  and excluded volume effects in a simple manner. By coupling it with the Kuramoto dynamics eq.~(\ref{eq:Kuramoto}), it allows us to study how these two ingredients, quantified by $\tau$ and $\phi$ respectively, affect the collective dynamical behaviour of a collection of coupled non-linear oscillators.

\section{Computational details}\label{sec:numerics}
In this work, we simulate the evolution of our model of interacting mobile oscillators over a broad range of parameters. We integrate eq.(\ref{eq:Kuramoto}) using an Euler scheme with step size $\Delta t$.
We fix $v_0=0.1$ and vary the persistence time in the range $\tau\in[0:10^3]$ by changing $v_1$. The time evolution is expressed in units of $\Delta t$ which is chosen small enough, such that the increment of $\theta_i$ at each step can not exceed $\pi/50$ and particle displacements can not exeed $v_0$ along each direction.  To explore finite size effects we simulate several system sizes from $N=1000$ up to $N=16000$. 
We vary the packing fraction from the ideal gas limit ($\phi=0$)  to $\phi=0.45$, well below jamming ($\phi_J\approx 0.80$). 

For systems with excluded volume  ($\phi>0$), studied in section~\ref{sec:disks}, we fix the diameter of the particles $\sigma=1$ and change the packing fraction by changing the linear size of the system $L$. 
We fix the probability that $N$ oscillators are within the interaction range to the value $\varphi=\frac{\pi R_{\theta}^2 \rho}{4}=0.49$ in order to compare systems at different densities but with the same local connectivity. While varying the density in the system, we adjust $R_{\theta}$  to conserve the mean number of neighbours. In this way we make sure that we analyse the impact of $\phi$ in the synchronization process.

For the systems of point particles studied in section~\ref{sec:points}, we fix the number density to $\rho=0.15625$. 
This choice has been made to allow the comparison between systems of point particles and ridig disks at the same number density. The corresponding packing fraction at this number density for a system of disks of diameter $\sigma=1$ is $\phi=0.12$ and the constraint $\varphi=0.49$ is fulfilled with $R_{\theta}=2$.
For point-like oscillators we simulate the model using different interaction radius, from local to long range  coupling.  
Note that the percolation threshold of the network generated by a homogeneous distribution of nodes at this density is at $R^p_{\theta}\approx 3$. This remark will become relevant in section~\ref{sec:regimes} where we discuss the deviations of our model from a limiting ideal case.

In order to study the synchronization dynamics, we let the system evolve from an completely disordered configuration of phases and follow how their tendency to align, together with their motion in  real space, gives rise to a collective coherent state.
We assign independently to each particle a random initial phase between $0$ and $2\pi$  from a uniform distribution.   
The initial positions we start with correspond to the steady states generated by the dynamics eq.~(\ref{eq:Motion01}).
Then we let the system evolve accordingly to eq.~(\ref{eq:Motion01}) and eq.~(\ref{eq:Kuramoto}).

The usual global order parameter measuring the degree of synchronization is defined by $Z(t)=N^{-1}|\sum_j e^{i\theta_j}|$~\cite{Kuramoto2012}. 
This quantity increases over time from its intial value $Z(0)\approx0$, characterising the initial incoherent state. If the system  synchronizes, $Z(t)$ asympotically reaches a steady value close to unity, meaning that all the oscillators have the same phase. 
A useful quantity to study the dynamics towards synchronization is the average phase difference defined as 
\begin{equation}
C_{\theta}(t)=\left\langle{ \sqrt{ \frac{2}{N(N-1)} \sum_{(i,j)} (\theta_i(t)- \theta_j(t))^2} } \right\rangle\, ,
\end{equation}
where the sum runs over all particle pairs and $\langle...\rangle$ denotes an average over several independent realisations of the dynamics.  
This function is expected to decay, after an initial transient, as an exponential with a characteristic time $T_s$, 
\begin{equation}
C_{\theta}(t)\sim e^{-t/T_s} \ .
\end{equation}
This  \emph{synchronization time} $T_s$ is a measure of the time needed to approach asymptotically a steady state of global synchronization characterised by $Z\approx1$. We will compute this quantity in our model over a broad range of parameteres and analyse under which conditions mobility induces a faster synchronization. 
As we describe in the following section, the behaviour of the synchronization time can be estimated from theoretical considerations in two limit regimes. Then, we will use these predictions as a  reference  when discussing our results. 

\section{Limit Dynamical Regimes}\label{sec:regimes}


For the sake of clarity, we should identify at this stage two extreme and opposite dynamical regimes that can be clearly identified. 
The first one corresponds to a regime where the topology of the network changes much faster that the phases of the oscillators. In this case, the so-called \emph{Fast Switching Approximation} (FSA) should be able to describe accurately the dynamics towards synchronization~\cite{Frasca2008,Fujiwara2011}. 
The opposite extreme case corresponds to a regime where phases change in a much shorter time scale than the links of the network.
We will call it \emph{Slow Switching} (SS) regime.   

\subsection{Fast Switching Regime}

Let us first consider the Fast Switching (FS) regime.
In network language, the FSA assumes that the time evolution of the links is much faster than the internal dynamics of the particles. In our case, this means that the motion of the agents, that induces the creation and deletion of links, is much faster than the phase changes. Then, in this limit, before there is a significative change in the internal dynamics, the links have been largely modified  and the  adjacency matrix can be replaced by its time average, where the entries correspond to the global probability that any two oscillators are within a distance smaller than $R_{\theta}$ under a completely random motion~\cite{Frasca2008, Fujiwara2011}.
Within this approximation, the synchronization time is given by
\begin{equation}\label{eq:FSA}
T^{\text{FSA}}_s=-\frac{1}{\ln(1-\pi K (N-1)/L^2)} \,,
\end{equation} 
which, to first order in the coupling strength, can be expressed as
\begin{equation}\label{eq:FSA}
T^{\text{FSA}}_s=\frac{L^2}{\pi K(N-1)}+O(K^2)\sim (\rho K)^{-1}\, .
\end{equation} 
Note that, in these expressions, the characteristics of the oscillators' motion (via $v_0$ and $\tau$) are absent. The mobility plays an indirect, yet crucial, role within this approximation. The velocity and persistence of the oscillators is related to the time scale characterizing the dynamics of the network, and then the validity of the approximation. The average  time needed for a self-propelled particle to diffuse over a distance of the order of the interaction radius is   $t_r=\pi R_{\theta}^2/D$, where $D$ is the diffusivity of the self-propelled agents. Another important time scale in the system is the persistence time $\tau$ which determines for how long particles moves ballistically.  These microscopic times should be compared with the time scale associated to the phase dynamics,  $t_{\theta}=K^{-1}$. 
Then, one expects that the FSA should hold when $t_{\theta}$ is larger than any other microscopic time scale, namely $t_{\theta}\gg \max[t_r,\,\tau]$.

In~\cite{Fujiwara2011}, one of the authors introduced a measure to explore quantitatively the regime of validity of the FSA
in a system made of Kuramoto oscillators carried by non-interacting Brownian particles far from the percolation thershold. The regimes we explore here do not fulfill this latter requirement (the percolation threshold is at $R_{\theta}\approx3$, see sec.~\ref{sec:numerics}) so the arguments of~\cite{Fujiwara2011} to establish a criteria that determines where we expect the FSA to hold, besides the large $t_{\theta}$ limit where FSA is always accurate, cannot be extended to our case. 

In order to understand quantitatively the origin of the deviations from the FSA behaviour in our model, we introduce a toy model which explicitly decouples the evolution of the network and the dynamics of the phases. In this Fast-Switching toy Model (FSM) the position of each particle is randomly reset at each time step, while keeping the Kuramoto dynamics untouched. In this way we implement numerically the limit conditions that corresponds to the assumptions of the FSA in our model. We  compare the results obtained with this toy model and our generic model of mobile agents in section~\ref{sec:points}, where we characterize the different dynamical regimes taking place in the system. 


\subsection{Slow Switching Regime}

Let us consider now the opposite extreme case  where the evolution of the network is so slow that it can be considered as static. 
In this limit, $t_{\theta}\ll\max[t_r,\,\tau]$, the oscillators are effectively immobile and the model defined by eq.~(\ref{eq:Kuramoto}) is equivalent to the  XY model with non-conserved order parameter dynamics (or Model A in the Halperin-Hohenberg's classification~\cite{Hohenberg1977}) at zero temperature. 
Indeed, eq.~(\ref{eq:Kuramoto})  can be rewritten as 
\begin{equation}
\dot{\theta}=-\frac{\delta}{\delta \theta}H\,, \ H=-\frac{K}{R_{\theta}^2}\sum_{i,j}  A_{ij}\, \boldsymbol{S}_i \cdot \boldsymbol{S}_j \, ,
\end{equation}
where we identified a spin with the phase of each oscillator $\boldsymbol{S}_i=(\cos\theta_i,\,\sin\theta_i)$. 
We recognise in this equation the Hamiltonian $H$ of the XY model in the network defined by $A_{ij}$. 
Note that the Kuramoto order parameter is equivalent to the absolute global magnetisation of the XY model: 
$Z(t)=\langle  \|\boldsymbol M(t) \| \rangle $ where $\boldsymbol{M}(t)=N^{-1}\sum_j \boldsymbol S_j(t)$.
Therefore, to study  the evolution towards synchronization of the Kuramoto model from an initially incoherent state is equivalent to  study  the relaxation dynamics of the XY model after a quench from $T\to \infty $ to $T=0$. 

For local connectivity, $R_\theta \ll L$, the interaction network is effectively two-dimensional.
It is well known from the Mermin-Wagner theorem that the 2$d$ XY model cannot display long-range order at any finite temperature. At the critical temperature $T_{KT}$, the system experiences a phase transition due to the unbinding of topological defects, the so-called Kosterlitz-Thouless phase transition~\cite{Kosterlitz1974}. In the high temperature phase the system is disordered, populated by free vortices.  Below $T_{KT}$ vortices bound into vortex-antivortex pairs and the system exhibits critical behaviour, meaning that the correlations decay algebraically.

The out-of-equilibrium dynamics of systems quenched through a symmetry breaking phase transition is a long standing problem in statistical mechanics (for a review in the subject see~\cite{Bray1994}).
The main reason being that it is among the rare  problems in statistical mechanics where general and precise theoretical statements can be made about systems evolving out-of-equilibrium. 
We briefly recall here the basic concepts related to this problem that we will use in our discussion about the synchronization of moving oscillators in section~\ref{sec:points} and~\ref{sec:disks}.

When a system in a homogeneous disorderd initial state - such as our incoherent initial state - is let to evolve towards an ordered state - such as the gobally synchronized state - it relaxes by locally growing ordered regions. The growth of these spatio-temporal heterogeneities in time is a \emph{coarsening} process.  
The central quantity that characterises the evolution of spatial phase structures  is the two-point correlation function
\begin{equation}
G(\boldsymbol{r},\,t)=\langle \boldsymbol{S}_i(t) \cdot \boldsymbol{S}_j(t)  \rangle_{\boldsymbol{r}_i-\boldsymbol{r}_j=\boldsymbol{r}} \, .
\end{equation}
The dynamic scaling hypothesis~\cite{Hohenberg1977, Bray1994} asserts that, in the long-time coarsening regime
\begin{equation}\label{eq:dynscaling}
G(r,\,t) \simeq \, F(r/\xi(t)) .
\end{equation}
Whether this scaling holds is a central issue in the study of phase ordering dynamics. It claims that, at late times, the dynamics following a quench  is characterised by a single growing length scale $\xi(t)$. 
Some approximation schemes have been established to obtain analytical expressions of the scaling function $F$. Among them, the Bray-Puri-Toyoki (BPT) approach, provides an expression of the scaling function for the case of  $n$-component vector models $O(n)$ with non-conserved dynamics~\cite{BrayPuri1991, Toyoki1992}.  
For the 2$d$ XY model (case $n=2$) quenched to $T<T_{KT}$~\cite{Yurke1993, Bray1994}
\begin{equation}\label{eq:scaling}
\xi(t,T)\sim \left[ \Lambda (T)\, \frac{t}{\ln t}\right]^{1/2}\, . 
\end{equation}
The logarithmic correction to the diffusive behaviour $\xi(t)\sim t^{1/2}$, expected for systems with non-conserved order parameter dynamics, is due to the presence of vortices. This scaling has been tested numerically in several works (for instance~\cite{Rojas1999}) and the parameter $\Lambda$ has been found to be an increasing function of temperature~\cite{Jelic2011}. 
In order to equilibrate, the system has to grow an equilibrium region of the order of the linear system size $\xi\sim L$. As a consequence, the relaxation time $t_{eq}$ needed to reach equilibrium grows with the system size. 
If we neglect the impact of logarithmic corrections in the equilibration time, the following scaling holds   
\begin{equation}\label{eq:FiniteSizeXY}
t_{eq}\sim L^{2}\ .
\end{equation} 
In the context of synchronization of locally coupled oscillators, this means that, in the limit of a static $2d$ network, the synchronization time  diverges as 
\begin{equation}\label{eq:FiniteSize}
T_s\sim N\ 
\end{equation}
up to logarithmic corrections. 


In the following, we analyse systematically the deviations from these two extreme regimes, FS and SS, in our model. We provide a comprehensive study of the conditions  under which  the synchronization of self-propelled oscillators follows the FSA predictions or, on the contrary, proceeds \textit{via} coarsening like in a static 2$d$ network.   

\section{Self-propelled point-like oscillators}\label{sec:points}

Before studying the role played by the heterogeneities due to the competition between self-propulsion and steric effects on the synchronization process, we focus first on a population of self-propelled point-like oscillators without excluded volume interactions ($\phi=0$). We  build our understanding about the impact of self-propulsion using this simplest situation to then  move to hard disk agents and address the effect of interactions in real space (see section~\ref{sec:disks}).

\begin{figure}[h]
\includegraphics[scale=0.75,angle=0]{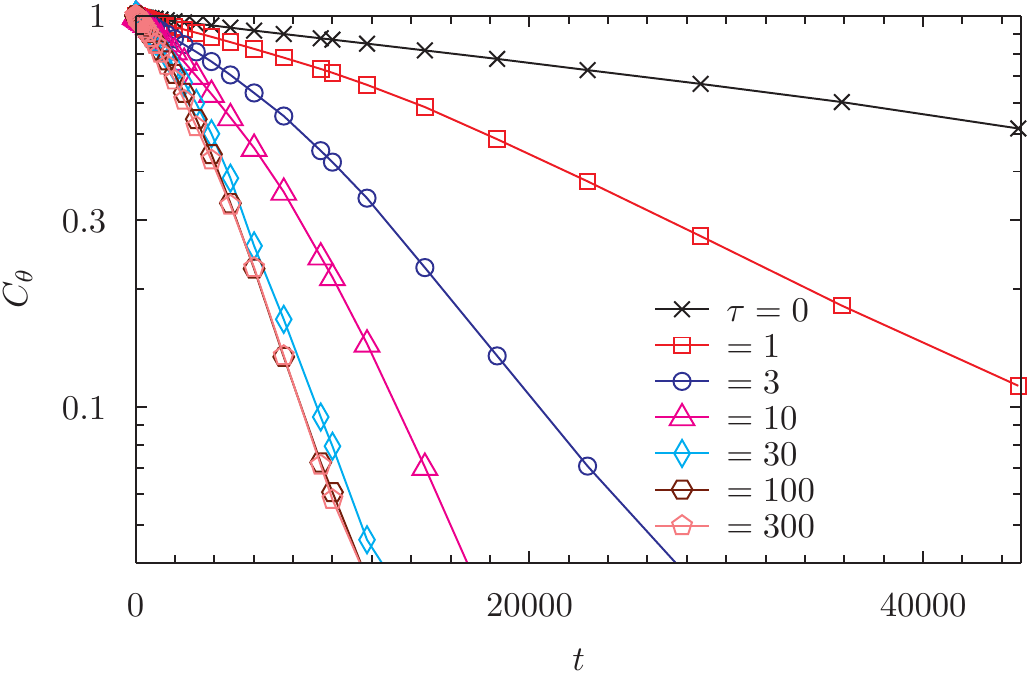}
\caption{(Color online) Evolution of the phase difference $C_{\theta}(t)/C_{\theta}(0)$ (after averaging over 500 independent runs) in lin-log scale for a system of $N=1000$ self-propelled point particles for $K=0.1$, $R_{\theta}=2$ and several values of  $\tau$.} \label{fig:C_t}
\end{figure}

\begin{figure}[h]
\includegraphics[scale=0.75,angle=0]{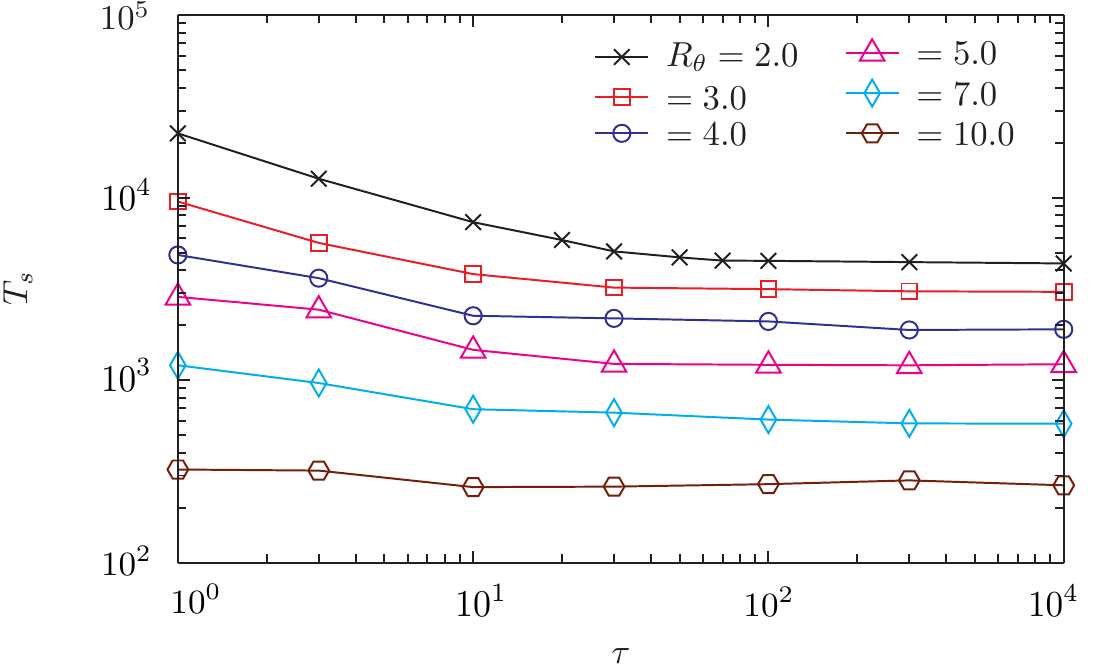}
\caption{(Color online) Synchronization time as a function of $\tau$ for several values of $R_{\theta}$ and $K=0.1$
} \label{fig:Ts_vs_tau_varR}
\end{figure}

In Fig.~\ref{fig:C_t} we show the evolution of $C_{\theta}$ for several values of $\tau$. As anticipated above, $C_{\theta}$ decays exponentially after an initial transient.  
The mean-square displacement quantifies how efficiently particles explore the available space, and therefore provides a measure of the degree of mixing in the system. 
Since self-propulsion reduces the mixing time, a given particle visits the neibourhood of other particles more often on average as $\tau$ is increased, accelerating the synchronization process. 
Similarly, the enhancement of synchronization by particle's motion has been previously reported using different models~\cite{Peruani2010, Uriu2012, Uriu2014}.
As shown in Fig.~\ref{fig:C_t} , the relaxation time $T_s$  decreases with $\tau$ up to a saturating value $\tau_{\text{sat}}$ above which $T_s$ is independent of $\tau$. 
Despite the fact that the diffusivity grows proportionally with the persistence time, $T_s$ is bounded for large values of $\tau$.
In order to understand the saturation of $T_s$ with $\tau$ one needs to consider the microscopic details of the synchronization process. 
%
As already pointed out, self-propelled particles motion display two different regimes, ballistic and diffusive.  
In the diffusive regime, two neighbouring agents interact, on average, for a period of time $t_r\approx \pi R_\theta^2/v_0^2\tau$: the characteristic time associated with the diffusion over the interaction area,  $\pi R_\theta^2$, associated to an oscillator. This time should be compared to the persistence time $\tau$. For $t_r<\tau$, the interaction between two oscillators occurs during the ballistic regime.
In this regime, the phase coupling proceeds between ballistic particles with identical velocity and the value of $\tau$, which sets the duration of such ballistic motion, does not play any role. The effect of different values of $\tau$ becomes relevant in the diffusive regime only. 
However,  the coupling between oscillators is not able to explore the dynamics of the particles in the diffusive time sector since particles are no longer whithin the interaction range when it sets in. 
Using this argument we find an estimation of the saturation threshold $\tau_{\text{sat}}\approx \sqrt{\pi} R_{\theta}/v_0$. For $R_{\theta}=2$ we get  $\tau_{\text{sat}}\approx 35.5$, in agreement with our data (see Fig.~\ref{fig:C_t} and Fig.~\ref{fig:Ts_vs_tau_varR}). 

From the exponential decay of $C_{\theta}$  we extract the synchronization time $T_s$ for several values of $\tau$ and $R_{\theta}$. The bare data is shown in Fig.~\ref{fig:Ts_vs_tau_varR}. For $\tau<\tau_{\text{sat}}$, $T_s$ is reduced by increasing self-propulsion.
In the saturated regime $\tau>\tau_{\text{sat}}$, $T_s$ has a value which only depends on $R_{\theta}$. As $R_{\theta}$ grows, the probability that two particles interact increases, thus reducing $T_s$. Note that the interacting network's connectivity increases with $R_{\theta}$ such that the impact of mobility, whose role is to connect originally distant oscillators for some period of time, is less pronounced. 
In the all-to-all limit, $R_{\theta}\to L$, all oscillators pairs are equivalent, so the synchronization process should be independent of their mobility. This is confirmed by our data which shows that the $\tau$-dependence of $T_s$ vanishes as $R_{\theta}$ is increased, approaching continuously the expected mean-field behaviour.

\begin{figure}[h]
\includegraphics[scale=0.70,angle=0]{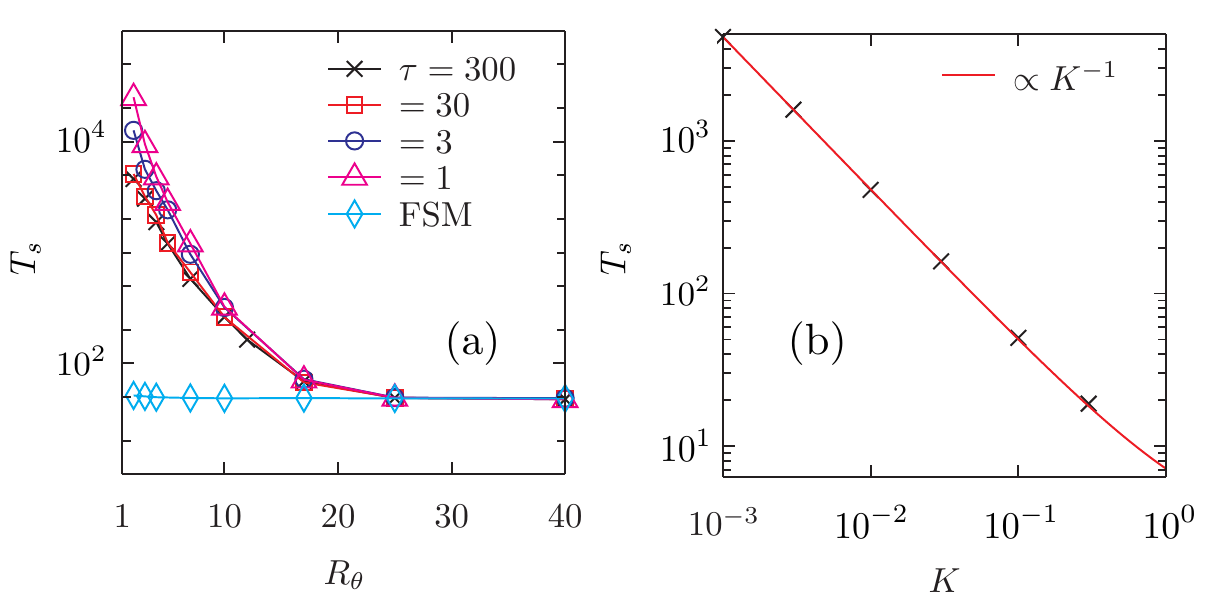}
\caption{(Color online) (a): Synchronization time as a function of $R_{\theta}$ for $K=0.1$ in a system of size $L=80$ with $N=1000$ particles. The horizontal points show the data obtained from our Fast-Switching toy model (FSM). (b): $T_s$ as a function of $K$ for our FSM with $R_{\theta}=2$. The solid red line indicates the FSA prediction eq.~(\ref{eq:FSA})} \label{fig:Ts_vs_R_vartau}
\end{figure}

As previously discussed,  the FSA and the all-to-all model share a crucial simplification: they both consider a fully connected interaction network and in this sense, they are both mean-field limits. The FSA justifies this simplification by a kinetic argument, while the mean-field Kuramoto model assumes that the coupling between all the oscillators is identical, independently of their location.    
In order to explore the connection between these two approximations, Fig.~\ref{fig:Ts_vs_R_vartau} (a) shows $T_s$ over a broad range of values of $R_{\theta}$. We compare the results obtained with simulations of the FSM, for which the topology of the network and the dynamics of the phases are decoupled.

As we vary the interaction radius from $R_{\theta}=2$ to $R_{\theta}=L/2$, $T_s$ decreases by three orders of magnitude. The data shows that $T_s$ decays as the all-to-all regime is approached, its asymptotic value being the one obtained from the FSM. We  have performed simulations of the latter toy model introduced in section~\ref{sec:regimes} for different values of $K$ at fixed $R_{\theta}=2$ (see Fig.~\ref{fig:Ts_vs_R_vartau} (b)). We have checked that, indeed, this simplified dynamics recovers the FSA behaviour. The data for $R_{\theta}\lesssim20$ clearly differs from the value of $T_s$ obtained with the FSM. As $R_{\theta}$ grows, the connectivity of the network increases until it is fully connected. Both extreme cases, FSA and all-to-all, give the same $T_s$, meaning that the topology of the network rules the synchronization process and that the FSA holds in the limit of $R_{\theta}\to L$. Moreover, the fully connected network and the FSA provides a lower bound for $T_s$ at a given value of $K$: synchronization cannot be achieved arbitrarily fast by increasing $R_{\theta}$ or $\tau$. 

\begin{figure}[h]
\vspace{0.2cm}
\includegraphics[scale=0.75,angle=0]{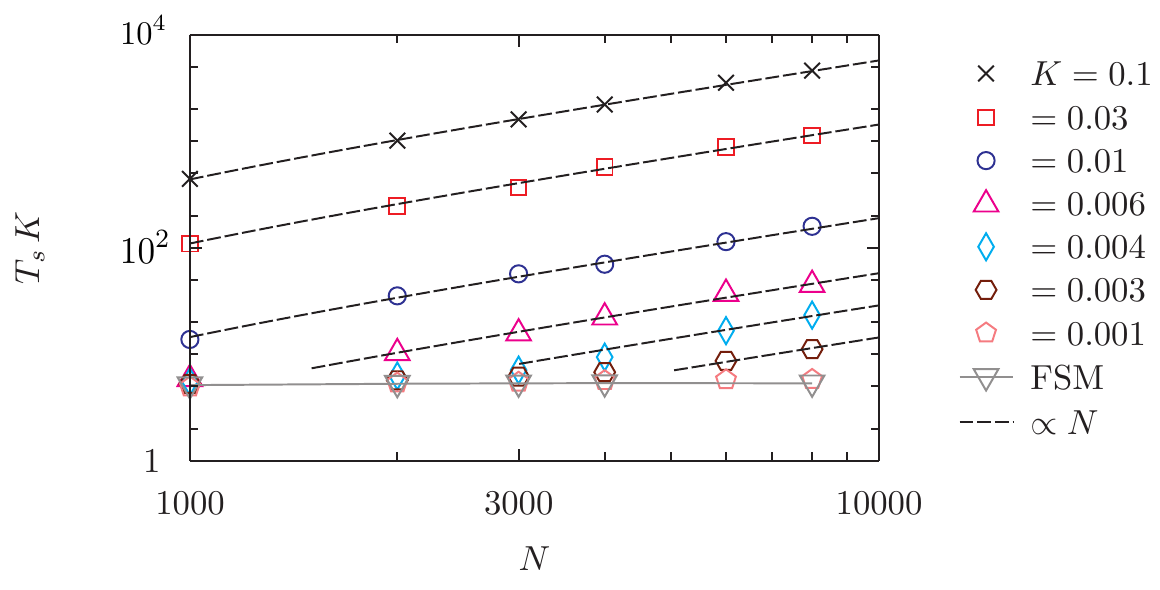} 
\caption{(Color online) Finite size behaviour of $T_s\,K$ for several values of the coupling strength $K$ shown in the key with fixed $\tau=300$ and $R_{\theta}=2$. The broken lines indicate the $T_s\sim N$ scaling expected in the SS regime. The FSM results are  shown in pointlines for comparison. } \label{fig:Ts_vs_N_varK}
\end{figure}

The approach to the FS regime can also be studied by increasing the natural time scale $t_{\theta}=K^{-1}$ of the Kuramoto oscillators. In the limit $K\to 0^+$ the dynamics of the phases and the network must decouple.
On the contrary, for large values of $K$ and $R_{\theta}\ll L$ we expect to approach the SS regime where, accordingly to eq.~(\ref{eq:FiniteSize}), $T_s\sim N$. 
In order to analyse further the approach to these two extreme dynamical regimes,
we explore the dependence of $T_s$ with the system size for a fixed short range coupling $R_{\theta}=2$. 
The data obtained for $\tau=300$ is shown in Fig.~\ref{fig:Ts_vs_N_varK}, together with  the simulation results of the FSM. For $K\geq 0.006$ the synchronization follows the   $T_s\sim N$ coarsening behaviour expected in the SS regime all over the explored range of system sizes. For smaller values of $K$ the range of validity of the coarsening scaling shortens and one needs to explore larger systems to eventually observe it. 
For $K\leq 0.006$, the synchronization time takes a constant value  - equal to the one obtained with the FSM - in small enough systems, while the coarsening process sets in at larger $N$.   
We observe a distinctive crossover from the FS behaviour $T_s\sim \text{constant}$ to the SS behaviour  $T_s\sim N$ as the system size is increased. 
As $K$ is reduced, the system approaches the FS regime and the range over which $T_s$ is constant grows. 
These results strongly suggest that, for any finite value of $K$,  the dynamical mechanism characterizing the evolution towards global synchronization in a system of locally coupled mobile oscillators is coarsening - for large enough system sizes above a threshold value that depends on $K$. We ought justify further this important claim in the following.  

\begin{figure}[h]
\vspace{0.2cm}
\includegraphics[scale=0.50,angle=0]{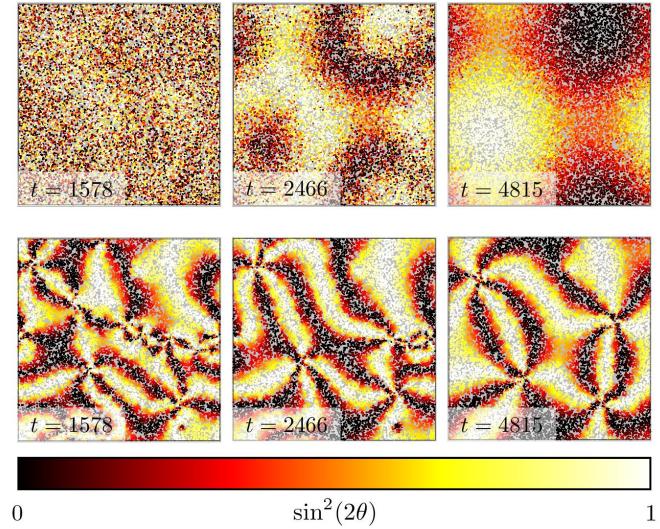} 
\caption{(Color online) Configurations at various times during the evolution from an incoherent initial state of two systems of size $N=16000$ with $\tau=300$, $K=0.001$ (top) and $K=0.1$ (bottom). The phase of the oscillators is represented as Schlieren patterns in which a color scale is associated to $\sin^2(2\theta)$. Each vortex emanates eight brushes of alternate black and white.} \label{fig:snap_d0}
\end{figure}

As shown in Fig.~\ref{fig:Ts_vs_N_varK}, the $T_s\sim N$  behaviour  is blurred by the system finite size, and this effect becomes more pronounced as $K$ decreases, approaching the FS regime. The snapshots shown in Fig.~\ref{fig:snap_d0} are in agreement with this picture. In the top row three configurations at $t=1578,$ $2466$ and $4815$ of a system with $\tau=300$ and $K=0.001$ are represented. Configurations at the same time and for the same value of $\tau$ are shown in the bottom row for a system with $K=0.1$. 
The emergence of time dependent phase heterogeneities is highlighted by the Schlieren patterns. This representation of the configurations is particularly adapted to visualize vortices: each vortex appears as a point from which four light and four dark brushes emanate, each color in between representing a different phase of the oscillators. Locally synchronized regions where particles share the same phase are identified as sections with the same color.   Schlieren patterns have been largely used to visualize systems with broken rotational symmetry, like vortices in the XY model~\cite{Yurke1993} or in nematic liquid crystals~\cite{Prost1995}. 
For $K=0.001$ the system develops large phase patterns that soon become of the order of the system size. We would need to simulate larger systems in order to be able to identify clearly the growth of a characteristic length scale  smaller than the linear size of the system. In this case, the growth of local order is arrested by the full homogenization of the system  and the scaling regime is not reached. 
For $K=0.1$ we observe a coarsening sequence reminiscent of the phase ordering dynamics of the 2$d$ XY model after a quench from $T\to\infty$ to $T<T_{KT}$. During the evolution, locally synchronized regions grow. The competition between degenerate coherent states with different average phase leaves behind topologically stable vortices. Then, the annihilation of oppositely charged vortices drives the dynamics at later times.

\begin{figure}[]
\includegraphics[scale=0.75,angle=0]{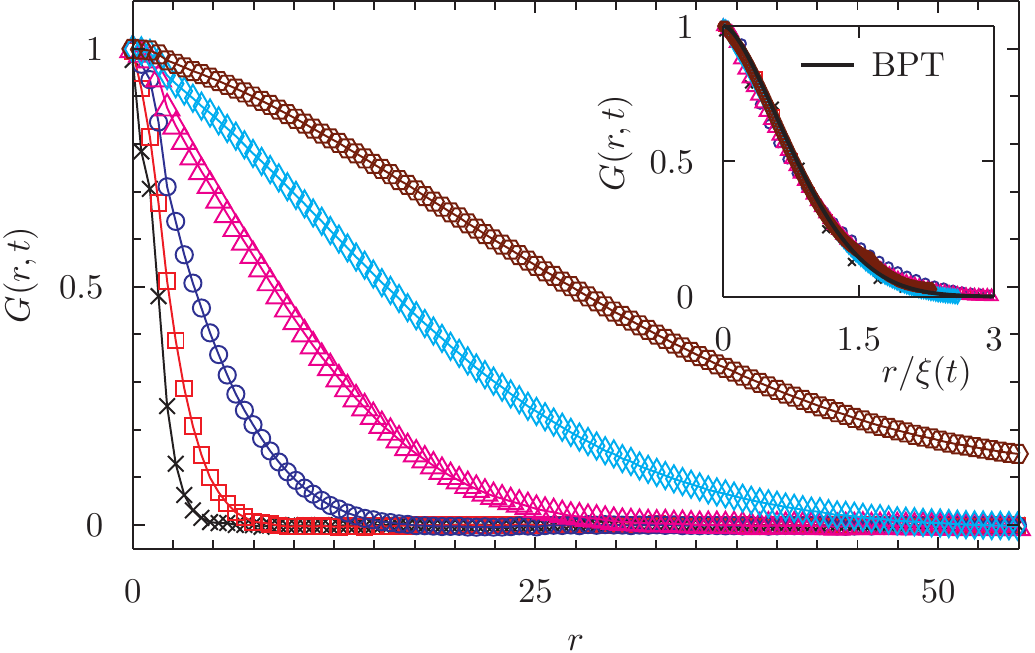} 
\caption{(Color online)  Space time correlation function $G(r,t)$ at different times $t=10, 29, 100, 331, 1000, 3082$ (from bottom to top) as a function of $r$. The data plotted is for a system of size $N=2000$ with $\tau=300$ and $K=0.1$. The inset shows the scaling plot $G(r,t)$ against $r/\xi(t)$ using the values of $\xi(t)$ extracted from  $G(\xi(t,\tau),t)=e^{-1}$. We show the BPT scaling function obtained after adjusting the continuous time variable in the analytical form with our numerical data.} \label{fig:DynSc_d0_G}
\end{figure}
\begin{figure}[]
\includegraphics[scale=0.80,angle=0]{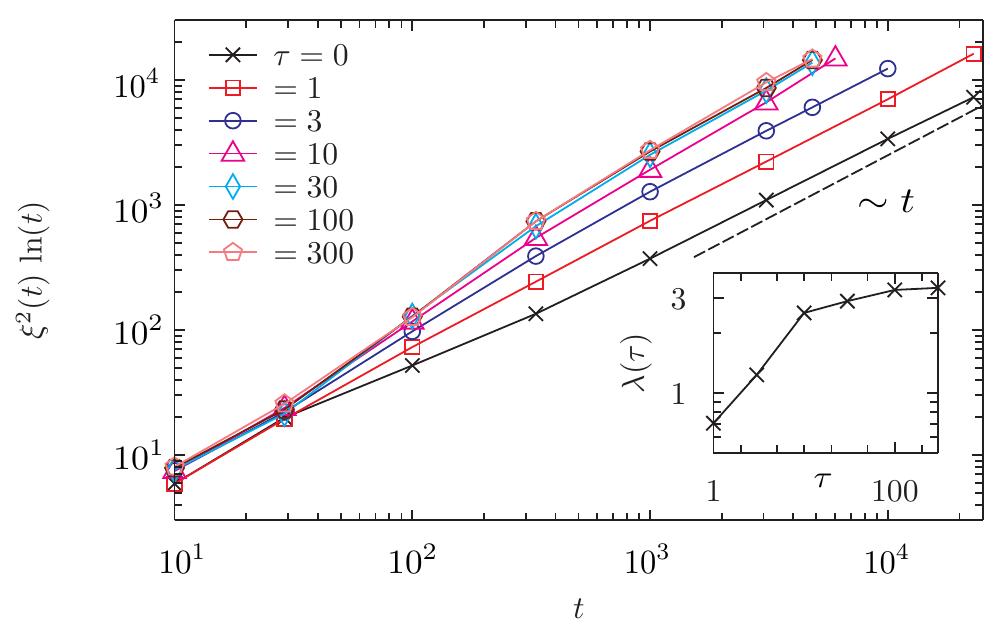} 
\caption{(Color online)  Growing length in log-log scale with respect to time for several values of $\tau$ given in the key. Inset: Dependence of the rate of growth $\lambda$ on $\tau$.} \label{fig:DynSc_d0_xi}
\end{figure}

A quantitative study of the growth of phase patterns 
can be achieved by computing the two point correlation function $G(r,t)$. 
In Fig.~\ref{fig:DynSc_d0_G} we show the decay of  $G(r,t)$ with respect to distance at several times for $\tau=300$ and $K=0.1$. 
We simulate systems of $N=2000$ particles, averaging over 300 independent runs for several values of $\tau$. We have checked that there are no significant finite-size effects in comparison to larger systems. 
We have extracted a characteristic length $\xi(t,\tau)$ from the decay of the correlations. The data obtained from the condition $G(\xi(t,\tau),t)=e^{-1}$ is depicted in Fig.~\ref{fig:DynSc_d0_xi}. 
As shown in the inset  Fig.~\ref{fig:DynSc_d0_G} the correlation function depends on space and time through the ratio $r/\xi$, thus verifiying dynamic scaling (see eq.~\ref{eq:dynscaling}). 
The scaling is excellent, and the BPT functional form  is in surprinsigly good agreement with the data. 

In Fig.~\ref{fig:DynSc_d0_xi} we plot $\xi^2(t)\ln t$ against time to compare the growth of $\xi$ in our model with the domain growth in  the XY model (see eq.~\ref{eq:scaling}).  
Spatial correlations develop faster as self-propulsion is increased for values of $\tau\lesssim \tau_{\text{sat}}\approx 35.5$.
Just as the phase difference $C_{\theta}$, the rate of growth of $\xi$ saturates for larger values of $\tau$ (see Fig.~\ref{fig:C_t}).  
After some initial transient, $\xi^2(t)\ln t$ grows linearly with time for all the values of $\tau$. 
We can extract the growing rate $\lambda$ from the slope of these curves.  
The values obtained from the best fits of the data are shown in the inset of Fig.~\ref{fig:DynSc_d0_xi}. 
Our simulations strongly suggest that the length scale characterising the size of phase synchronized regions grows asymptotically in time as
\begin{equation}
 \xi(t,\tau)\sim \left[ \lambda (\tau)\, \frac{t}{\ln t}\right]^{1/2}\ .
\end{equation}
Note the similarity between this equation and eq.~\ref{eq:scaling} describing the relaxation of the 2$d$ XY model. The non-universal prefactor $\Lambda(T)$ has been replaced by $\lambda(\tau)$ here, which depends on the persistence time instead of temperature. 
The  small logarithmic deviations to the $\xi\sim t^{1/2}$ diffusive law can hardly be measured using the data shown in Fig.~\ref{fig:DynSc_d0_xi} (more decades in time would be needed to produce reliable data in this respect).   

In the XY model, the temperature dependence of the growing length is captured by  $\Lambda(T)$. Numerical simulations have found that this prefactor increases linearly with temperature~\cite{Jelic2011}, but a complete theoretical understanding of this behaviour is still laking. In the absence of a thermal bath, the only source of noise in our model is the non-Markovian stochastic motion of the agents, quantified by the persistence time $\tau$.
In this point-like limit, the diffusivity of the particles $D\propto v_0^2\tau$ which, from a generalized Stokes-Einstein relation, can be interpreted as an effective temperature $T_{\text{eff}}\propto \tau$~\cite{Levis2015, Ginot2015}. 
As shown in Fig.~\ref{fig:DynSc_d0_xi}, $\lambda$ is an increasing function of $\tau$, hence $T_{\text{eff}}$, up to the saturation value $\tau_{\text{sat}}$. 
Despite that our model displays the same dynamical universal features as the 2$d$ XY model, it is reasonable to expect that the non universal prefactor $\lambda$ depends on the specific dynamics of the system.    
A similar dependence on the persistence time has also beeen observed for the synchronization time 
(see Fig.~\ref{fig:Ts_vs_tau_varR}). This provides further justification to our claim that the dynamical mechanism ruling the relaxation towards a globally synchronized state is the one of growing synchronized domains separated by topological defects. The increase and saturation of  $\lambda$ with $\tau$ has the same origin as the $\tau$-dependence of the synchronization time $T_s$. These  two quantities display a similar dependence on the parameters of the system and are therefore said to be coupled.   


In this section, we have indentified and characterised quantitatively the different dynamical mechanisms taking place during the synchronization of a collection of phase oscillators carried by non-interacting self-propelled agents.
We have shown that self-propulsion accelerates the synchronization process up to a threshold value  determined by the local connectivity of the interaction network.
We studied the impact of the connectivity scheme by varying the interaction radius. In the limit of a fully connected network, the FSA becomes exact and the mobility mechanism of the particles irrelevant. 
In this regime, all the particles in the system synchronize at the same rate, which means that the mechanism leading the dynamics is \emph{global}. 
This regime where the FSA is accurate can also be reached in locally connected networks for small enough values of $K$. 
Above this limit of small coupling, the synchronization mechanism of locally coupled mobile oscillators is \emph{local}. In this latter regime, the system exhibits a coarsening process analogous to the phase ordering dynamics of the 2$d$ XY model following a quench. 
Despite the fact that the underlying network structure is mobile and that the dynamics breaks detailed balance, the evolution of our model verifies  dynamic scaling, like a system in contact with a thermal bath.  

%


\section{Self-propelled hard disks}\label{sec:disks}

We turn now into finite packing fractions, where many-body effects may impact the synchronization of locally coupled mobile agents. We explore systems at several values of $\phi$ with fixed  particle diameter $\sigma=1$ and coupling $K=0.1$. Since we want to identify the $\phi$ and $\tau$ dependence of the synchronization process, we fix the probability that $N$ oscillators are within the interaction range to the value $\varphi=0.49$ (see sec.~\ref{sec:numerics}).
This choice  corresponds to the short range  coupling regime ($R_{\theta}\ll L$).

\begin{figure}[]
\includegraphics[scale=0.4,angle=0]{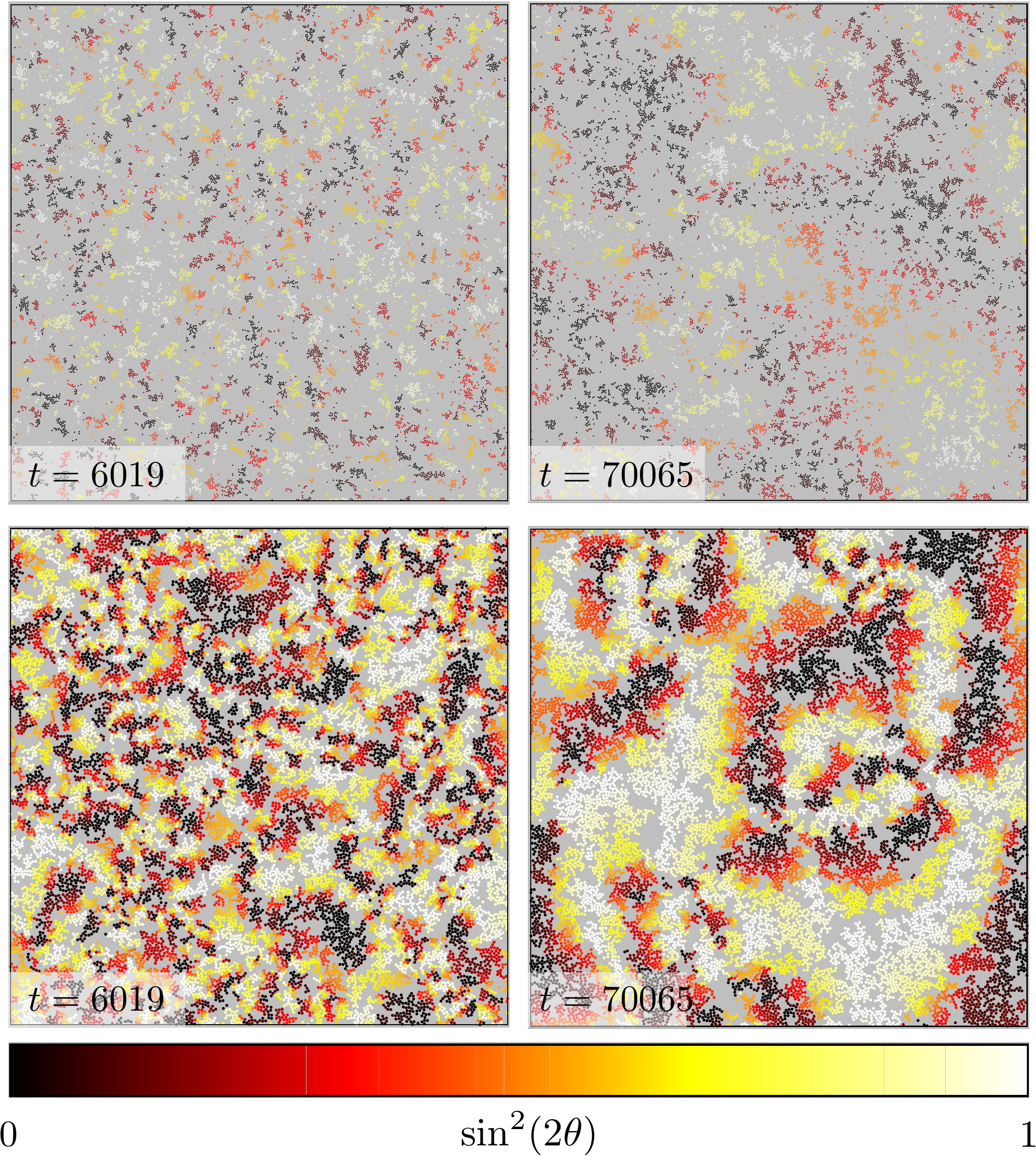} 
\caption{(Color online) Snapshots of a system made of $N=16000$ self-propelled hard disks at two different times $t=6019$ and $70065$ with $\tau=300$, $\phi=0.12$ (top) and $\phi=0.45$ (bottom).  As in Fig.~\ref{fig:snap_d0}, the phase of each particle is represented by a color scale of the amplitude of $\sin^2(2\theta)$.} \label{fig:snap_d1}
\end{figure}
 

Before studying the synchronization process, it is useful to briefly describe the structural properties of the underlying active fluid of self-propelled hard disks,
since the liquid structure dictates the nature of the dynamic network where the oscillators evolve. 
Adding excluded volume interactions between self-propelled particles leads to the emergence of complex non-equilibrium spatial structures in the system~\cite{Levis2014}. Particles have the tendency to aggregate, even in the absence of any attractive interaction, forming clusters that grow as $\tau$ is increased.
This is a genuine out-of-equilibrium effect due to the competition between self-propulsion and excluded volume interactions. 
Two typical steady state configurations of the system at two different packing fractions are shown in Fig.~\ref{fig:snap_d1}. 
By steady state we mean here an asymptotic state obtained from the motion of the particles in  real space, independently of the dynamics of their internal oscillators. 
Once a steady state has been reached, the Kuramoto dynamics is turned on from such `liquid steady state' and the evolution towards a synchronized state governed by eqs.~(\ref{eq:Motion01}),~(\ref{eq:Motion02}) and (\ref{eq:Kuramoto}) is studied. 
In Fig.~\ref{fig:snap_d1} we  illustrate this evolution by showing two configurations at different times starting  from a random distribution of phases. 

\begin{figure}[]
\includegraphics[scale=0.75,angle=0]{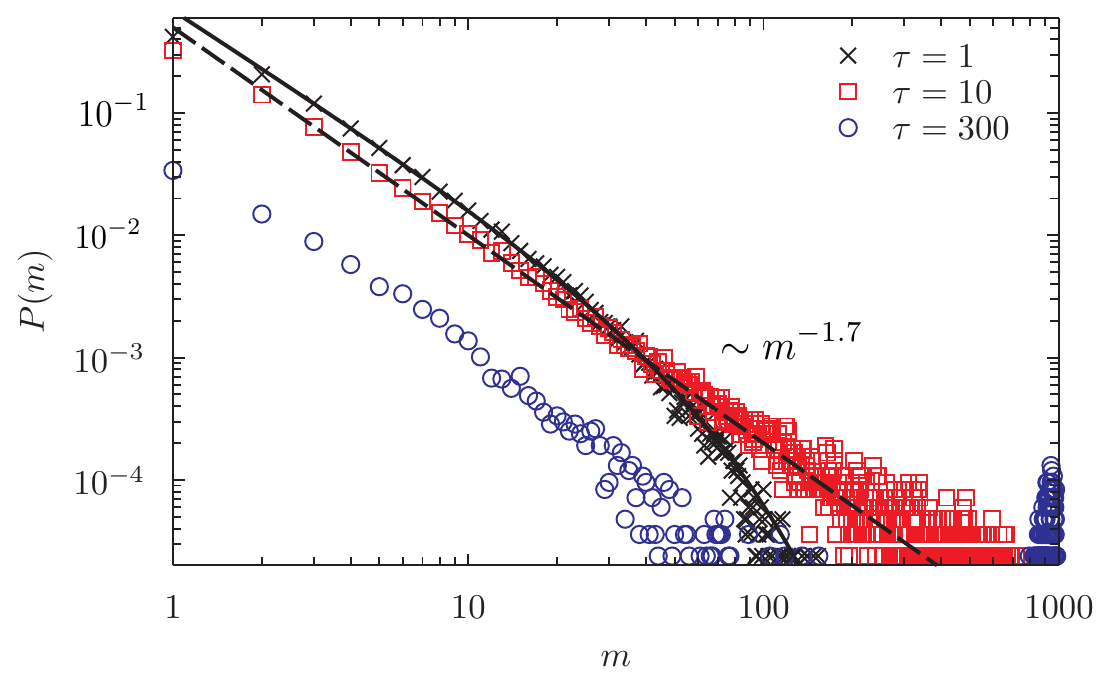}
\caption{(Color online) Cluster size distribution for a system composed by $N=1000$ disks at $\phi=0.45$ and several values of $\tau$. The percolation threshold in this system is at $\tau_c\approx 10$. The continuous line represents the distribution eq.~\ref{eq:Pm}. The broken line corresponds to the power law behaviour $P(m)\sim m^{-\nu}$ at percolation with $\nu=1.7$.
} \label{fig:Pm}
\end{figure}

A cluster is defined by a set of connected particles, meaning that the distance between them is smaller than $R_{\theta}$ so the oscillators they carry are coupled. 
We characterize the cluster structure in the steady state by means of the cluster size distribution function $P(m)$,  defined as the probability to find a particle in a cluster of size $m$.  
In Fig.~\ref{fig:Pm} we show the evolution of $P$ at  $\phi=0.45$ and increasing persistence. 
As discussed in detail in~\cite{Levis2014}, the distribution broadens as $\tau$ is increased, meaning that larger clusters are formed. Since the cluster size increases with $\tau$, for  high enough self-propulsion, a macroscopic cluster spanning the whole system can eventually emerge. 
The formation of a percolating cluster for $\tau$ above a critical value $\tau_c$ is analog to the formation of a gel-like structure in suspensions of attractive colloids. Indeed, the persistence time in this model plays the role of an effective attraction~\cite{Ginot2015}. 

Below the percolation threshold, the numerical data is well reproduced by the following functional form of the distribution
\begin{equation}\label{eq:Pm}
 P(m)=\frac{e^{-m/m^*}}{m^{\nu}}\, ,
\end{equation}
where $\nu\approx 1.7$ and $m^*\approx65$. The same functional form has been found in previous instances of particle aggregation induced by activity~\cite{Peruani2012,Soto2014,Levis2014}. 
At percolation, the distribution becomes algebraic $P(m)\sim m^{-\nu}$, meaning that clusters of any size  populate the system. For $\tau=\tau_c$ the system is critical. 
Above the percolation point, $\tau>\tau_c $, macroscopic clusters have a non-negligible weight, represented in the distribution by a peak at large cluster sizes $m\approx N$ (see Fig.~\ref{fig:Pm}).    

\begin{figure}[]
\includegraphics[scale=0.65,angle=0]{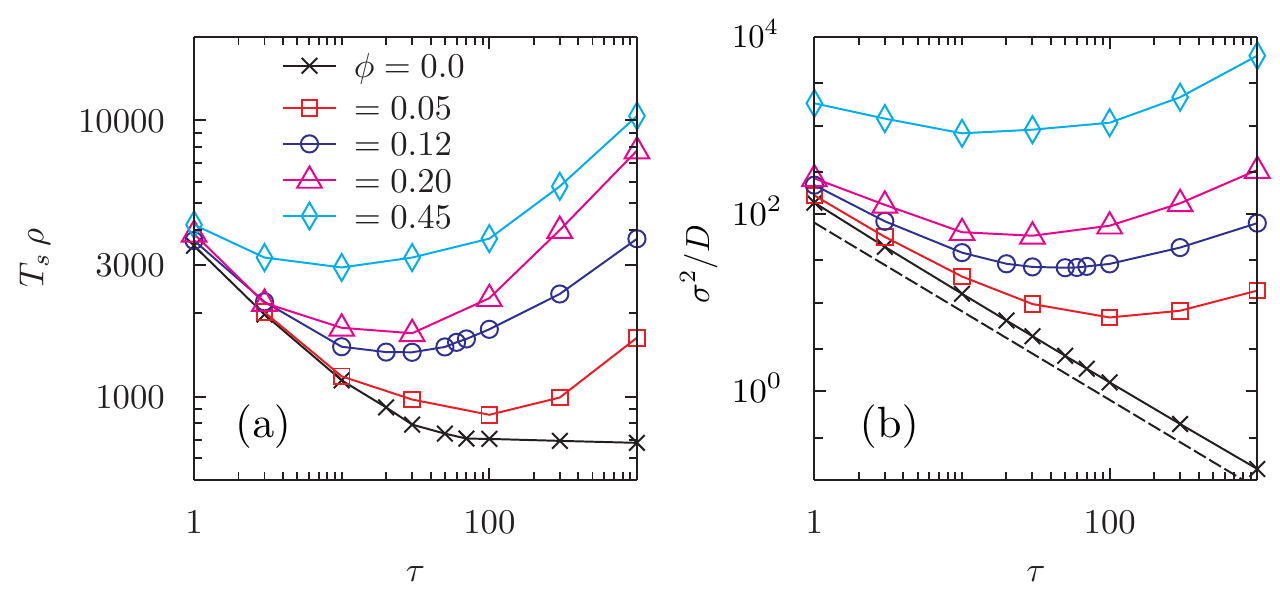}
\caption{(Color online) (a): Synchronization time $T_s$ times the number density $\rho$ as a function of $\tau$ for several values of $\phi$ shown in the key. 
We present the results of $T_s$ normalized by $1/\rho$ since $1/\rho K$ introduces a time scale that can be absorbed in the time units. 
(b): Diffusion time for the same set of parameters as the ones in the left panel. The broken line represents the dilute limit behaviour $\sigma^2/D\propto \tau^{-1}$.  The data for $\phi=0$ with $R_{\theta}=2$ and $\rho=0.15625$ is also shown as a reference. 
} \label{fig:Tsync_d1}
\end{figure}

We move now into the analysis of the synchronization time and study the impact of the heterogeneous structures described above. In Fig.~\ref{fig:Tsync_d1} we show the results of extensive numerical simulations for $T_s$ and $D$ over a broad range of values of $\tau$ and $\phi$. 
As already mentioned in section~\ref{sec:points} and confirmed by our data, the diffusion coefficient $D$ in the limit $\phi\to 0$ is proportional to the persistence time. 
However, many-body effects strongly alter this simple behaviour. The diffusion coefficient at finite density is a non-monotonic function of the persistence time~\cite{Levis2014}. As shown in Fig.~\ref{fig:Tsync_d1} (b), there is a density-dependent optimum value of the persistence time $\tilde{\tau}(\phi)$ for which the diffusion time $\sigma^2/D$ is minimal.
The emergence of this optimum value is a direct consequence of particle clustering. At small values of $\phi$ and $\tau$,  small clusters coexist with very dilute regions where particle collisions are rare. Particles in these large voids move basically free,  such that, as for non-interacting particles, the diffusivity increases proportionally with $\tau$. For larger values of $\phi$ and $\tau$, bigger clusters are formed, leaving smaller voids where particles can move faster.  The competition between these two opposite effects leads to the observed non-monotonic behaviour of $D$. Increasing  self-propulsion accelerates the particles but also leads to the formation of increasingly large clusters where particles inside are kinetically trapped. Since clusters are more likely to appear at larger packing fractions,  $\tilde{\tau}(\phi)$ decreases with $\phi$.

As  shown in Fig.~\ref{fig:Tsync_d1} (a), the synchronization time displays an analogous non-monotonic behaviour. 
There is an optimal value of $\tau$ that minimizes the time needed to reach global synchronization for a given packing fraction. Moreover, this value is identical to the one that minimizes the diffusion time, namely $\tilde{\tau}(\phi)$. 
This observation, is in agreement with our previous discussion about the synchronization of point-like agents. Since the  diffusion  gives a characteristic time for particle mixing, one expects $T_s$ and  $D$ to be higly correlated.  
The structure of the active liquid affects the synchronization time through the presence of clusters. The existence of a percolation threshold does not  have a direct impact on $T_s$. The important structural feature having a direct impact on the synchronization is the growth of  clusters  with $\tau$ and $\phi$. 
At a given number density $\rho$, synchronization is slower at higher packing fractions. The saturated regime at high persistence time, found in the ideal gas system, is never reached in the range of packing fractions explored here.


We now investigate to what extent the presence of excluded volume interaction alters the dynamical mechanisms we discussed in the previous section. In particular,  the asymptotic behaviour of the dynamical correlation length and the similarities between the synchronization process and the dynamic scaling of the $2d$ XY model. We follow then the same line of reasoning as above, but now, for a system where self-propulsion and excluded volume interactions compete.

From the four configurations that illustrate the evolution of the phase of the self-propelled disks (see Fig.~\ref{fig:snap_d1}), one clearly observes the growth of synchronized (or ordered) domains. 
The dynamics  proceeds first through the local synchronization within a cluster. The presence of steric effects results in the formation of clusters, so particles inside have a natural preference to synchronize their phases. 
As the evolution proceeds, synchronized regions of typical size $\xi(t,\tau)$ grow. 
The competition between synchronized regions with different phase leads to the formation of vortices. 
We have observed the same behaviour for point-like agents. 
The presence of particle interactions does not seem to modify the stability of topological defects.   
At later times, once the characteristic length has become larger that the mean cluster size, the system settles into a much slower dynamical regime.
This regime is characterized by the annihilation of oppositely charged vortices in a similar way to what happens in the 2$d$ XY model and the simplified version of our model discussed in the previous section.

\begin{figure}[]
\includegraphics[scale=0.75,angle=0]{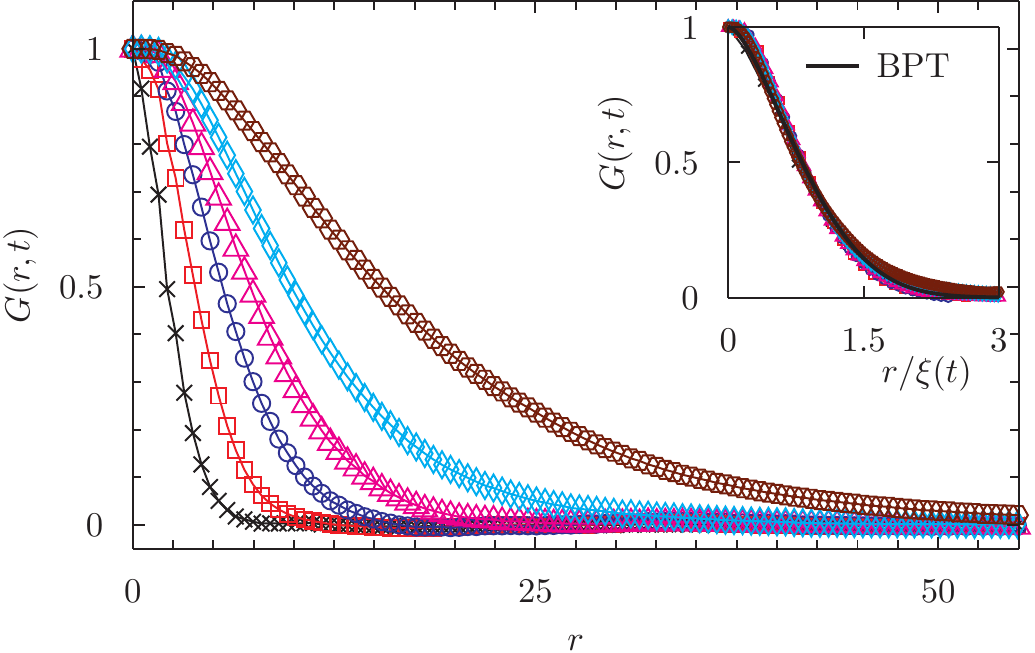}
\caption{(Color online) Space time correlation function $G(r,t)$ at different times $t=10$, $29$, $100$, $331$, $1000$, $3082$ (from bottom to top) for a system of size $N=2000$ with $\phi=0.12$, $\tau=300$ and $K=0.1$. 
The inset shows the scaling plot where distances have been normalized by the characteristic length $\xi(t)$ extracted from  $G(\xi(t,\tau),t)=e^{-1}$. We show the BPT scaling function in the time units that better fit our numerical data.} \label{fig:DynSc_d1}
\end{figure}

\begin{figure}[]
\includegraphics[scale=0.75,angle=0]{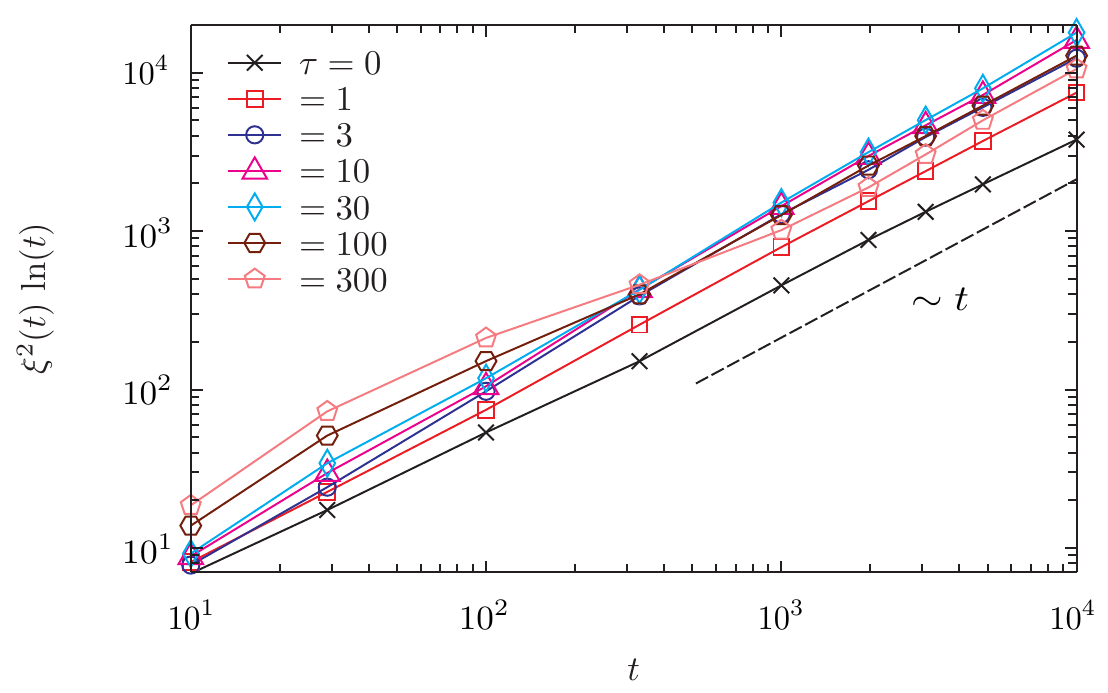}
\caption{(Color online) Time dependence of the growing correlation length in a system of size $N=2000$ for $\phi=0.12$ and several values of $\tau$ shown in the key. We also represent for comparison   a linear growth $\sim t$ in dashed lines. } \label{fig:xi_d1}
\end{figure}

The behaviour of the space time correlation function $G(r,t)$ confirms this growth. As shown in Fig.~\ref{fig:DynSc_d1} space correlations grow in time.  
Consistently with the data shown in  Fig.~\ref{fig:Tsync_d1}, synchronization takes longer to establish in a system at finite packing fraction, as it is visible by comparing Fig.~\ref{fig:DynSc_d1} with  Fig.~\ref{fig:DynSc_d0_G}. 
In the regime where domains grow and $\xi \ll L$, the  correlation function $G $ depends on space and time through $\xi(t,\tau)$ only.
The data shown in the inset Fig.~\ref{fig:DynSc_d1} confirms this claim. 
Notably, the master curve in the inset is well reproduced by the BPT functional form. 

In order to characterize further the growth of order in the system we compute $\xi$ over a broad range of values of $\tau$ using $G(\xi(t,\tau),t)=e^{-1}$. The data is represented in Fig.~\ref{fig:xi_d1} as $\xi^2 \ln(t)$ against $t$ in log-log scale. 
In the long time regime, we find that $\xi^2 \ln(t)$ grows linearly in time. 
Our data suggests that $\xi$ verifies the scaling form eq.~\ref{eq:scaling}.
At short times $t<\tau$, synchronized domains grow faster when $\tau$ is increased. This corresponds to the short time dynamical regime described above where particles adjust their phases inside the  clusters. Since clusters grow monotonically with $\tau$, the characteristic length $\xi$ does as well in this regime.
At later times, the $\tau$-dependence of $\xi$ is no longer monotonic. 
In this second, later time regime,  particles have to adjust their phases at larger scales  in order to reach global synchronization. This leads to a slower dynamics where the  structure of the underlying active liquid plays a non-trivial role.

\begin{figure}[]
\includegraphics[scale=0.75,angle=0]{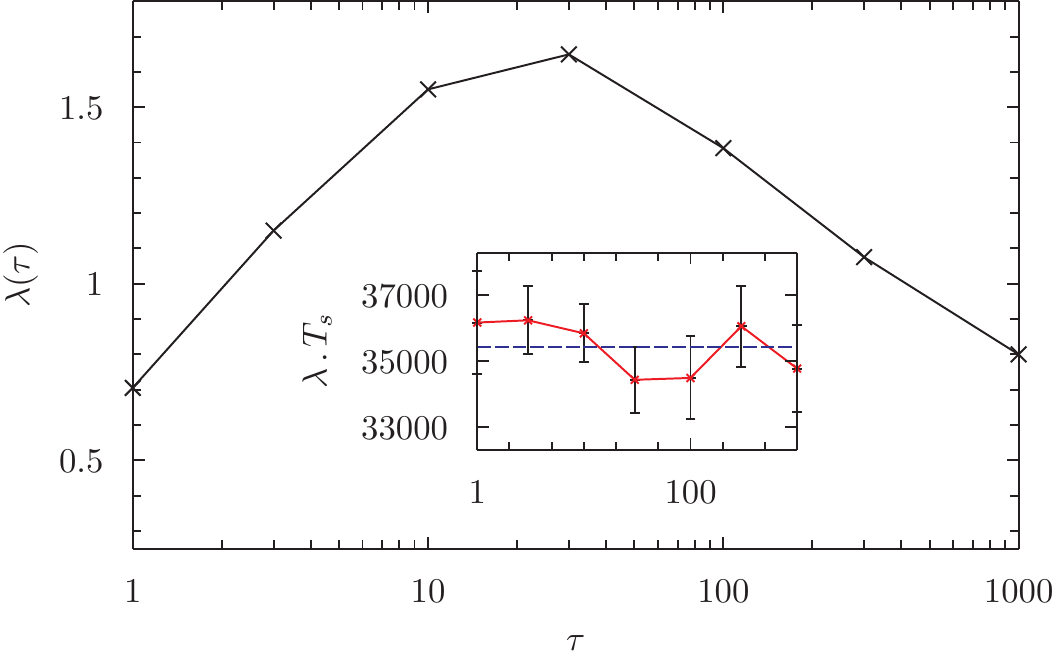}
\caption{(Color online) Dependence of $\lambda$  with $\tau$. In the inset we show that $\lambda\,  T_s \approx 35451$ with a reasonable degree of accuracy of 1\%.} \label{fig:lambda_d1}
\end{figure}

A direct measure of the $\tau$-dependence of the growth process is provided by $\lambda$, the slope of the plots  $\xi^2 \ln(t)$ against $t$. The results are shown in  Fig.~\ref{fig:lambda_d1}. 
As already discussed, for values of $\tau<\tilde{\tau}$ particle mixing is more efficient when $\tau$ is increased, explaining the rise of $\lambda$ al low persistence times. At higher values of $\tau$, the formation of large clusters reduces the ability of the particles to mix across the whole system and then to globally synchronize, leading to the drop of $\lambda$. This behaviour is consistent with the non-monotonic dependence of $T_s$ with $\tau$ already discussed. The crossover time between
 these two regimes is given by $\tilde{\tau}$, the same optimal value for which $T_s$ and $D^{-1}$ are minimal. 
This suggests that $\lambda$ and $T_s$ are strongly coupled, that is, both measures provide the same information about the long time synchronization dynamics. Indeed, as shown in Fig.~\ref{fig:lambda_d1}, $\lambda \times T_s \approx \text{constant}$ up to numerical accuracy, meaning that the dynamical process leading the approach to a synchronized state is coarsening, and that the rate of growth of locally ordered regions determines the synchronization time.

In this section, we have investigated how the presence of interactions between agents in  real physical space affects the synchronization of mobile oscillators. The main impact of excluded volume interactions is the emergence of non-equilibrium clusters at finite packing fraction, which gives rise to a non-monotonic dependence of the synchronization time with the persistence time. The mechanism by which locally coupled phase oscillators synchronize in this heterogeneous, time-evolving network, is coarsening, and the domain growth characterizing the dynamics is consistent with the dynamic scaling behaviour of the 2$d$ XY model with non-conserved order parameter dynamics.

\section{Conclusions}\label{sec:conclusion}

We have introduced a simple model of synchronization in time-evolving networks, where the nodes are interacting physical self-propelled objects. We have found that, in the absence of particle-particle interactions, self-propulsion promotes synchronization of locally coupled oscillators. Interestingly, in the presence of repulsive interactions, synchronization can be optimized by choosing a precise value of the persistence time for a given density. Such an optimum comes from a delicate balance between the enhancement of particle motion and the tendency to form clusters as we increase the persistence of the particles, a purely out-of-equilibrium many-body effect. 
This new effect shows that the interplay between the oscillator coupling and the topology of the underlying network, arising from particle interactions, plays a crucial role for the performance of mobile systems which might be seen as an evolutionary factor in living systems.
We have explored the range of validity of two opposite limit regimes where the synchronization process can be well understood from theoretical arguments. 
The first regime corresponds to the limit where the motion of the particles and the dynamics of the phases can be considered as being decoupled and the synchronization time can be computed analytically using the so-called Fast-Switching Approximation. 
The second regime corresponds to the limit case where the motion of the particles is much slower that the dynamics of the phases and the synchronization dynamics is equivalent to the phase ordering dynamics of the 2$d$ XY model after a quench. 
We have shown that the Fast Switching approximation holds in two limits: (i) in the limit where the range of phase coupling is large, and hence the motion is irrelevant; (ii) in the limit of very weak coupling for small enough systems, since in this case the exchange of  neighbours is a fast process compared with the phase interaction time scale.
Besides these limit situations, synchronization of locally coupled moving oscillators generically proceeds through coarsening. Despite its non-thermal nature, the model fulfills the dynamic scaling hypothesis and the same scaling laws as the 2$d$ XY seem to hold for any coupling strength in the limit of large system size. Our results support the idea that our model belongs to the dynamical universality class of the 2$d$ XY model with Model A dynamics and that the evolution of the network where the model is defined does not alter its scaling behaviour. In this way, we make a fundamental connection between the dynamics of a non-equilibrium active system and thermal one which fulfills the fluctuation-dissipation theorem. 

The new model proposed in this work  sheds light into the generic mechanisms leading the synchronization of mobile physical entities. Although concrete experimental systems, like bacteria colonies or robot swarms, might have their own specificities (not precisely captured by our model), the general features of  synchronization we have identified will be at play for any system where motility, synchronization and excluded volume coexist. 
The insight obtained constitutes a useful guideline to address the nature of synchronization in systems where the internal degree of freedom and the particle motility are coupled, and help designing optimizing synchronization strategies for artificial autonomous agents. 

Here we have considered the internal phases of the particles as being completely decoupled from their motion. A natural extension of the model that would allow to study the emergence of collective motion in active systems, would be to couple the internal phase with the direction of motion. 
As such, our results should be useful as a guideline for the design of artificial autonomous agents, and in particular for  the optimization of their synchronization strategies. 
Understanding the impact that motility and excluded volume have in the onset of synchronization for populations of agents with a distribution of natural frequencies persists as a relevant challenge. It remains unclear whether the competition between the tendency of mobile oscillators to order in the dynamical network topology and the intrinsic frequency randomness can hinder synchronization. 
In future work we aim at investigating these different scenarios where the nature of the steady-state itself has not been elucidated yet.

\section*{Acknowledgements}

D. L. acknowledges funding from the European Union's Horizon 2020 research and innovation programme under the Marie Sk\l odowska-Curie Actions/(H2020-MSCA-IF) Grant Agreement No. 657517.
I. P. acknowledges support from MINECO (Spain), Project FIS2015-67837-P, DURSI Project 2014SGR-922, and Generalitat de Catalunya under Program ICREA Acad\`emia. A.D.-G. acknowledges financial support from MINECO, Projects FIS2012-38266 and FIS2015-71582, and from Generalitat de Catalunya Project 2014SGR-608.

\bibliographystyle{apsrev}
\bibliography{synchro.bib}{}

\end{document}